\documentstyle[sprocl]{article}

\input{epsf}
\def\Journal#1#2#3#4{{#1} {\bf #2}, #3 (#4)}

\def\NPB{{\em Nucl. Phys.} B}
\def\PLB{{\em Phys. Lett.}  B}
\def\PRL{\em Phys. Rev. Lett.}
\def\PRD{{\em Phys. Rev.} D}
\def\PRB{{\em Phys. Rev.} B}

\def\be{\begin{equation}}
\def\ee{\end{equation}}
\def\bea{\begin{eqnarray}}
\def\eea{\end{eqnarray}}
\begin{document}
\title{PROPERTIES OF QCD VACUUM FROM LATTICE}

\author{ A. DI GIACOMO}

\address{Pisa University, 2 Piazza Torricelli,
Pisa,\\ 56100 Pisa, ITALY}
\maketitle\abstracts{
Advances in the understanding of the basic 
properties of QCD vacuum will be reported. Three main subjects will be touched:
\begin{itemize}
\item[1)] Condensation of monopoles and confinement.
\item[2)] Topology, or instanton physics.
\item[3)] Gauge invariant field strength correlators, and their behaviour
across the deconfining phase transition.
\end{itemize}
}
\section{Introduction}
I will report on some progress recently achieved mainly by numerical
simulations on the lattice, on three subjects. 1) Condensation of
monopoles and colour confinement (sect.2). 2) Topology (instanton physics)
(sect.3). 3) Gauge invariant field strength correlators at short distances at
$T=0$ and at the deconfining transition (sect.4).

The conclusions of this review will be that:
\begin{itemize}
\item[1)] Solid evidence exists that dual superconductivity~\cite{1,2,3} is the
mechanism of colour confinement.\cite{8,9}
\item[2)] The Witten-Veneziano~\cite{4,5} formula for the mass of the $\eta'$ is
definitely correct.
This statement has been made possible by the construction of an improved
(almost perfect) operator for the topological charge density.\cite{6}
By use of
the same technical improvement  a detailed study of the topological
susceptibility across the deconfining transition has been made, showing a sharp
drop of it at
$T_c$.\cite{7}

Studies with dynamical fermions have been started, especially with the aim of
determining, in full QCD, $\chi$, its derivative $\chi'$ with respect to momentum
transfer, and the matrix element of topological charge on proton states, which
would allow a measurement from first principles of the so called ``spin content
of the proton''. All these results in full QCD have been slowed down by the
discovery~\cite{6bis} that the usual algorithm for Montecarlo simulation, the so
called hybrid montecarlo, is very slow in changing the topological charge $Q$ of
a configuration: the same $Q$ stays for few hundred updatings. This puts
severe limitations on the validity of usual montecarlo studies in full QCD.
A typical statistics is indeed a few hundred configurations. Only for
observables which are insensitive to topology can such a statistics be
sufficient to insure proper thermalization.
\item[3)]
Previous studies~\cite{29} of the field strength correlators in the vacuum have
been extended to much shorter distances ($\sim 1 $~fm). The results is relevant
to test stochastic models of confinement.
\end{itemize}
All these results have been made possible by the use of  QUADRIX computers.
\section{Condensation of monopoles and confinement.}
An appealing mechanism for colour confinement in QCD is dual superconductivity
of the vacuum. The chromoelectric field acting between heavy $q$~$\bar q$ pairs
is constrained into Abrikosov flux tubes, generating a potential energy
proportional to the distance. 

We test this mechanism from first principles by numerical simulation on the
lattice.\cite{8,9}

The basic idea of the approach is that, to detect dual superconductivity of QCD
vacuum, the vacuum expectation value of an operator with non trivial magnetic
charge can be used as a probe, or as disorder parameter. Indeed a non zero
value of such {\em v.e.v.} signals condensation of monopoles and spontaneous
breaking of the $U(1)$ symmetry related to magnetic charge conservation. The
relevant magnetic $U(1)$'s are identified by the well known procedure of abelian
projection.\cite{10} They correspond to residual $U(1)$ invariance after
diagonalization of any operator belonging to the adjoint representation of the
gauge group. Each operator in the adjoint representation defines an abelian
projection, and hence a species of monopoles.

We have constructed a creation operator for monopole, and we use its {\em
v.e.v.} as a disorder parameter.

It is an open question if all abelian projections are equivalent, and identify
dual superconductors.\cite{10} We have tested the abelian projection defined by
the diagonalization of the Polyakov line, which is the local operator defined as
the parallel transport on the closed path along the time axis, starting from a
point and coming back to it through the periodic boundary conditions used to
define finite temperature. The {\em v.e.v.} of the trace of such operator is
the usual order parameter for confinement. We find definite evidence for dual
superconductivity in the $U(1)$ defined by such abelian projection.\cite{9}

The creation operator for an $U(1)$ monopole is defined as follows~\cite{8}:
\begin{equation}
\mu(\vec y,t) = {\rm exp}\left[
{\rm i}\int {\rm d}^3 x\,\vec E(\vec x,t)\frac{1}{e}\vec b(\vec x - \vec
y)\right]
\label{eq:1}\end{equation}
$\vec E$ is the electric field operator, and
\[ \vec b(\vec r) = \frac{m}{2}\frac{\vec r\wedge\vec n}{r(r - \vec r\cdot \vec
n)}\]
is the vector potential describing the field of a monopole of magnetic charge
$m$ in units $1/2e$. The prescription for the integral is that the Dirac string
must be removed. $\mu$ is the analog of the translation operator for  a
particle:
\[{\rm e}^{{\rm i} p a}|x\rangle = |x+a\rangle\]
$\vec E$ is the conjugate
momentum to the field $\vec A$. In the Schr\"odinger representation
\begin{equation}
\mu(\vec y,t)|\vec A(\vec x,t)\rangle =
|\vec A(\vec x,t) + \frac{1}{e}\vec b(\vec x - \vec
y)\rangle\label{eq:2}\end{equation}
$\mu$ adds a monopole to the field configuration. Moreover, if
$Q_M = \int{\rm d}^3 x\,\vec\nabla(\vec\nabla\wedge\vec A)$ is the magnetic
charge
\begin{equation}
\left[Q_M,\mu(\vec y,t)\right] = m\,\mu(\vec y,t)\label{eq:3}\end{equation}
i.e. $\mu$ carries magnetic charge $m$.

The lattice version of $\mu$ is
\begin{equation}
\mu =  {\rm exp} -\left(\beta S_M\right)
\label{eq:4}\end{equation}
with
\begin{equation}
S_M = \sum_{\vec n} \Pi_{0i}(\vec n,n_0)\left({\rm e}^{{\rm i} b_i} - 1\right)
\label{eq:5}\end{equation}
$\frac{1}{e}{\rm Im} \Pi_{0i}$ is the lattice version of the electric field. 
The disorder parameter is
\begin{equation}
\langle \mu\rangle = \frac{\displaystyle
\int {\rm d} U\,{\rm exp}\left[-\beta(S_W + S_M)\right]}
{\displaystyle
\int {\rm d} U\,{\rm exp}\left(-\beta S_W \right)}
\label{eq:4ter}
\end{equation}

Eq.~\ref{eq:5} corresponds to Eq.~\ref{eq:1} at the lowest order in $\vec b$,
but it is a compactified version of it. 
By use of the definition Eq.~\ref{eq:5} the gauge arbitrariness in the choice
of $\vec b$ to describe the monopole field is reabsorbed in the Haar measure of
the Feynman integral. To avoid problems with fluctuations we do
not measure
$\langle\mu\rangle$ directly, but $\rho = \frac{{\rm d}}{{\rm d}\beta}\ln\mu$
and reconstruct $\langle\mu\rangle$ as
\begin{equation}
\langle\mu\rangle = {\rm exp}\int_0^\beta\rho(x)\,{\rm
d}x\label{eq:6}\end{equation}
$\langle\mu\rangle$ determined in this way converges to a nonzero constant as
$V\to\infty$ for $T < T_c$ and goes to zero as $(T_c-T)^\delta$ at the
deconfining temperature. The typical form of $\rho$ is shown in fig.~1 for
$SU(3)$ on a $12^3\times4$ lattice. The sharp negative peak at $T_c$ 
signals the drop of $\langle\mu\rangle$ at $T_c$. $\langle\mu\rangle$ is
zero for $T>T_c$ only in the limit $V\to \infty$, as is typical of disorder
parameters.\cite{8,11}
\vskip0.05in
{\centerline{
\epsfxsize0.85\linewidth
\epsfbox{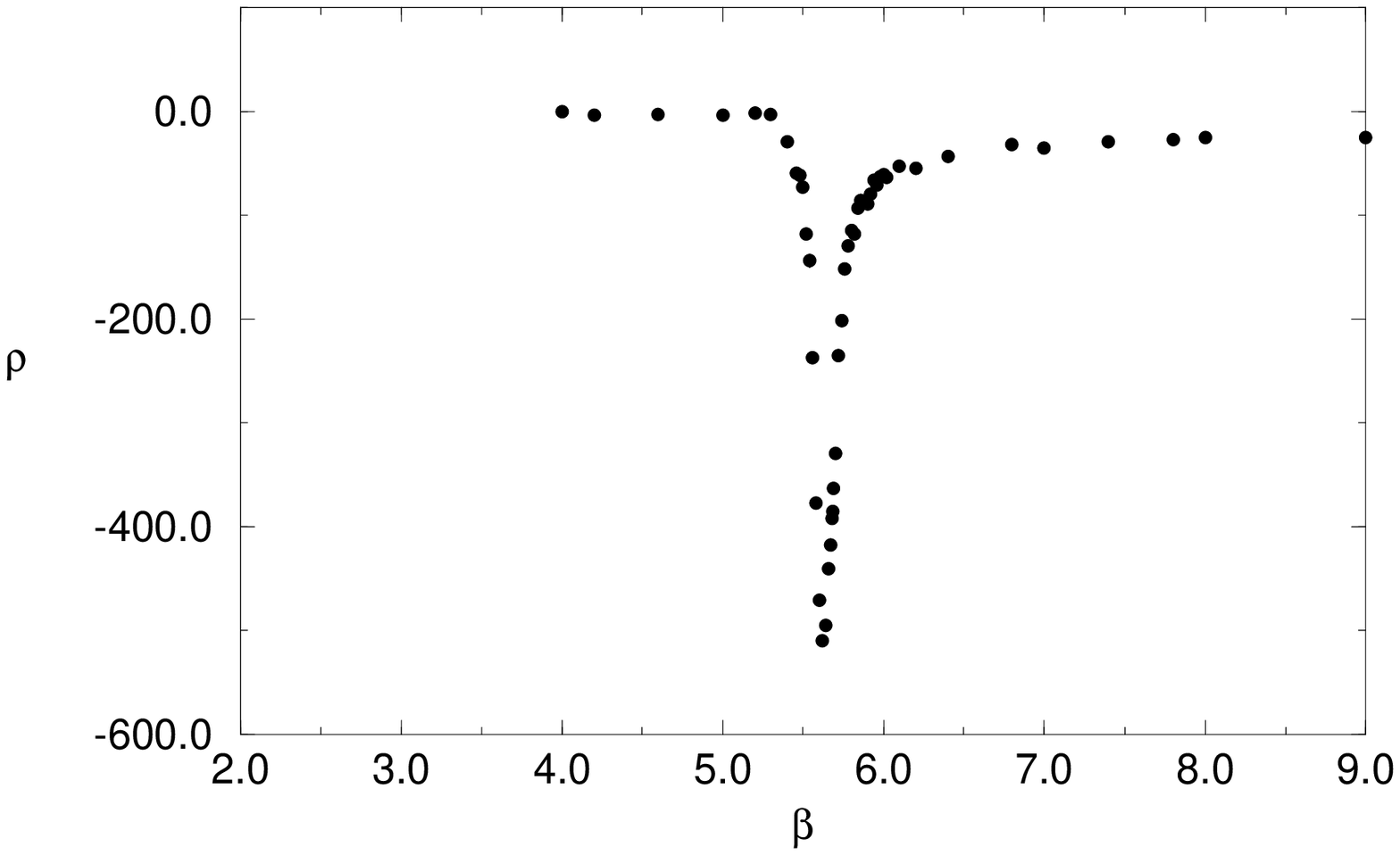}
}}
\vskip0.05in
{\centerline{{\bf Fig.1}\,\,
$SU(3)$ gauge theory. Lattice $12^3\times 4$. $\rho=\frac{{\rm d}}{{\rm
d}\beta}\ln\langle\mu\rangle$}}
\vskip0.1in
We have repeated~\cite{12} the construction 
for the $XY$ model in 3-d,
both as a test of the method and as an amusing
application of it. The model describes superfluid
$He_4$. It can be viewed as the euclidean version  of
a free massless particle in $2+1$ dimension.

The field variable is an angle $\theta(x)$. On the lattice:
\begin{equation}
S = \beta\sum_{\mu,i}\left[1 - \cos(\Delta_\mu\theta)\right]
\mathop\simeq_{a\to
0}\frac{\beta}{2}(\Delta_\mu\theta)^2\label{eq:7}\end{equation}
The theory admits soliton configurations which have the topology of
vortices
\begin{equation} 
\bar\theta(\vec x -\vec y) = {\rm arctg}\frac{\displaystyle (\vec x-\vec y)_2}
{\displaystyle (\vec x-\vec y)_1}\label{eq:8}\end{equation}
is an example. Vortices condense in the ground state in the superfluid phase:
they play a similar role as monopoles in QCD. As for QCD we can define a
creation operator for a vortex
\begin{equation}
\mu(\vec y,t) =
{\rm exp}\left[{\rm i}\int\partial_0\theta(\vec x,t)\bar\theta(\vec x-\vec
y)\,{\rm d}^2 x\right] \label{eq:9}
\end{equation}
The operator which counts vortices, $V$ can be written as
\begin{equation}
V = \frac{1}{2\pi}\int{\rm d}^2 x\,\left(\vec\nabla\wedge\vec A\right)\qquad
\vec A = \vec \nabla\theta\label{eq:10}\end{equation}
and
\begin{equation}
\left[V,\mu\right] = \mu\label{eq:11}\end{equation}
On the lattice by the same procedure of compactification used for monopoles
\begin{equation}
\mu(\vec y,x_0)=
{\rm exp}\left\{-\beta\left[\sum_{n\neq 0}\cos\left[\Delta_0\theta(\vec n) +
\bar\theta(\vec n - \vec y)\right]
-\cos\left[\Delta_0\theta(\vec n)\right]\right] \right\}
\label{eq:12}\end{equation}
As in the case of monopoles we define $\langle\mu\rangle$ in terms of the
correlator of a vortex antivortex by cluster property
\begin{equation}
\lim_{|x|\to\infty} \langle\mu(x)\mu(0)\rangle \simeq
A{\rm e}^{-\alpha |x|} + \langle\mu\rangle^2\label{eq:13}\end{equation}
and instead of $\langle\mu\rangle$ we measure
\begin{equation}
\rho = \frac{{\rm d}}{{\rm d}\beta}\ln\langle\mu
\rangle\label{eq:14}\end{equation}
The behaviour of $\rho$ is shown in Fig.~2.

A finite size scaling analysis, described in
ref.\cite{12} gives, for the critical index of the correlation length
\begin{equation}
\nu = 0.669 \pm 0.065 \qquad (0.670(7))
\label{eq:15}\end{equation}
for the critical temperature
\begin{equation}
\beta_c = 0.4538 \pm 0.0003 \qquad(0.45419(2))\label{eq:16}\end{equation}
and for the critical index of $\langle\mu\rangle$
\begin{equation}
\delta = 0.740\pm0.029\label{eq:17}\end{equation}
The first two quantities agree with the determinations by other methods, which
are shown in parentheses, showing that the method is correct and effective.
\vskip0.05in
{\centerline{
\epsfxsize0.85\linewidth
\epsfbox{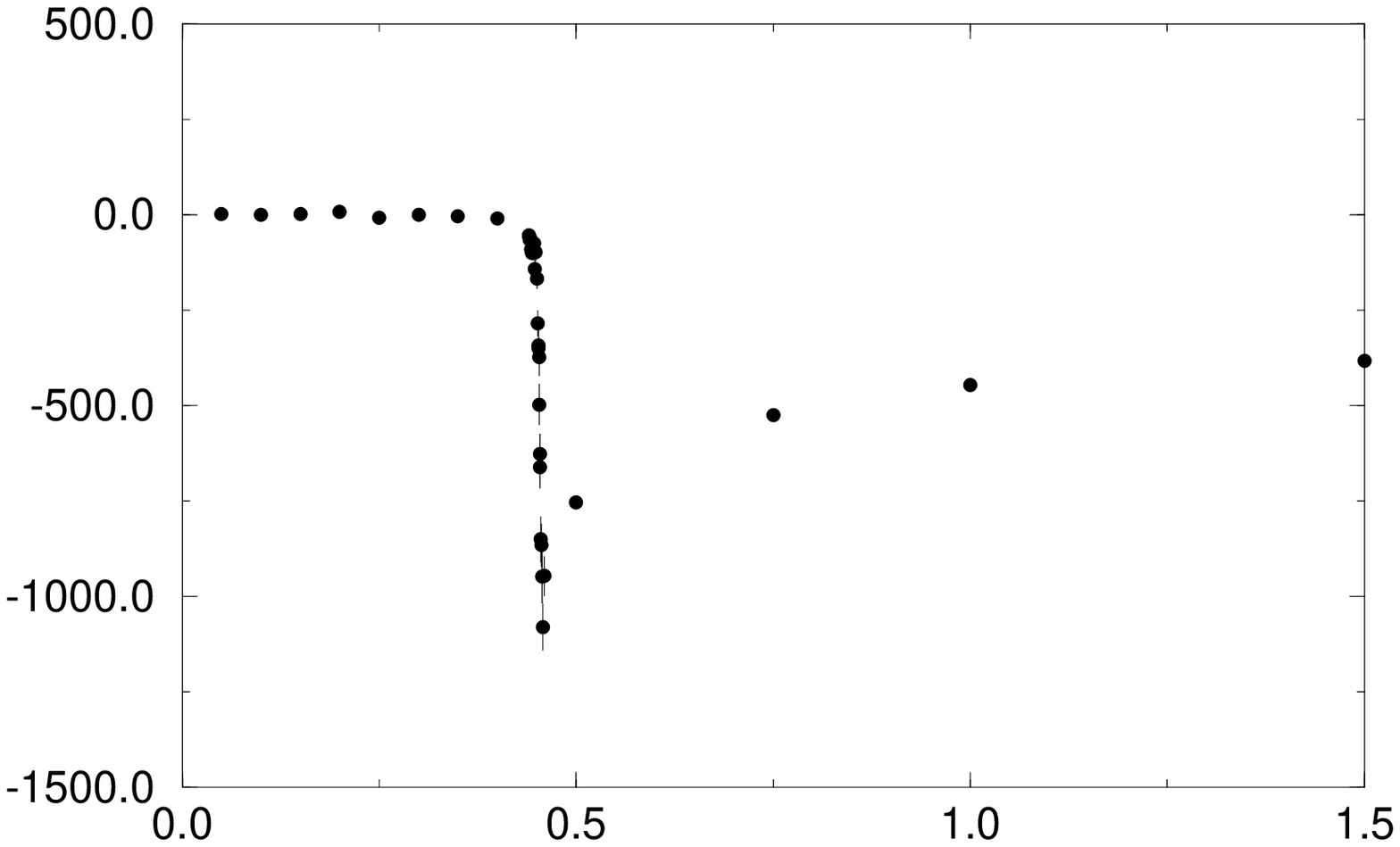}
}}
\vskip0.05in
{\centerline{{\bf Fig.2}\,\,
$XY$ model. Lattice $40^3$. $\rho=\frac{{\rm d}}{{\rm
d}\beta}\ln\langle\mu\rangle$ v.s. $\beta$.}}
\vskip0.1in
\section{Topology (instantons)}
Almost all of the Gell-Mann current algebra, which was originally abstracted
from massless free quark model, has been preserved with the advent of QCD,
except for the singlet axial current, $j_5^\mu = \sum_f
\bar\psi_f\gamma^\mu\gamma^5\psi_f$, which was conserved in the quark model,
but is anomalous in QCD
\begin{equation}
\partial_\mu j_5^\mu = 2 N_f Q\label{eq:18}\end{equation}
$Q(x) = \frac{g^2}{64\pi^2}\varepsilon^{\mu\nu\rho\sigma}G^a_{\mu\nu}
G^a_{\rho\sigma}$ is the density of topological charge.\cite{13} $Q = \int{\rm
d}^4 x\,Q(x)$ is the second Chern number, is known as topological charge and
takes on integer values on smooth configurations. The non consevation of
$j^5_\mu$ in QCD gives a handle to solve the so called $U(1)$ problem: if
$j^5_\mu$ were conserved and the corresponding symmetry were a Wigner symmetry,
parity doublets should exist, if it were realized \`a la Goldstone then one
should have $m_{\eta'} \leq\sqrt{3} m_\pi$ and neither of these possibilities is
realized in nature.\cite{14}

However in an expansion in $1/N_c$ $Q(x)$ is non leading, being $\propto g^2
=\lambda/N_c$ ($\lambda = g^2 N_c)$. In a philosophy in which the leading order
in $1/N_c$ describes the essentials of hadron physics,\cite{15,16} $U(1)$ can be
considered a simmetry. The anomaly acts there as a perturbation and displaces
the $\eta'$ mass from zero, which would correspond to Goldstone symmetry, to its
true value.\cite{17,18} The quantitative relation is~\cite{18}
\begin{equation}
\frac{2 N_f}{f_\pi^2}\chi = m_\eta^2 + m_{\eta'}^2 - 2\,m_K^2\qquad{\rm or}\;
\chi = (180\,{\rm MeV})^4
\label{eq:19}\end{equation}
$\chi$ is the topological susceptibility of the vacuum
\begin{equation}
\chi = \int{\rm d}^4 x\,\langle 0| T\left(Q(x) Q(0)\right)|0\rangle
\label{eq:20}\end{equation}
at the leading order, i.e. in the absence of fermions.

The prediction Eq.~\ref{eq:19} can be tested on lattice and involves the
computation of $\chi$ in the quenched approximation. The correct way to compute
$\chi$ is to define a regularized version of $Q$, $Q_L$, and compute the lattice
topological susceptibility
\begin{equation}
\chi_L = \sum_n\langle  Q_L(n) Q_L(0) \rangle \label{eq:21}\end{equation}
Like in any regularization scheme the regularized operator mixes, in the limit
in which the cutoff is removed, with all the operators having the same quantum
numbers and smaller or equal dimension in mass.\cite{19,20}

By use of this prescription one gets
\begin{eqnarray}
Q_L &=& Z(\beta) Q \label{eq:22a}\\
\chi_L &=& Z^2(\beta)\chi a^4 + B(\beta) G_2 a^4 + P(\beta)\label{eq:22b}
\end{eqnarray}
In Eq.~\ref{eq:22b} $G_2 =
\langle\frac{\alpha_s}{\pi}G^a_{\mu\nu}G^a_{\mu\nu}\rangle$ is the gluon
condensate, and $P(\beta)$ describes the mixing to the identity operator, which
is usually called perturbative tail. Both the last two terms in
Eq.~\ref{eq:22b} come from the singularity in the definition (\ref{eq:21})
of $\chi$ as
$x\to 0$. $Z(\beta)$, $B(\beta)$, $P(\beta)$ depend on the choice of $Q_L$, which is
arbitrary for terms of ${\cal O}(a^6)$ or higher. For the most simple choice of
$Q_L$, in terms of the plaquette $\Pi^{\mu\nu}$
\begin{equation}
Q_L = -\frac{1}{32\pi^2}\varepsilon_{\mu\nu\rho\sigma}{\rm
Tr}\left\{\Pi^{\mu\nu}(n)\Pi^{\rho\sigma}(n)\right\}\label{eq:23}\end{equation}
$Z=0.18$ and the correction to continuum in Eq.~\ref{eq:22b}, namely the sum of
the last two terms in Eq.~\ref{eq:22b} is much bigger than the first term.

A nonperturbative determination both of $Z$ and of the additive renormalization
$M(\beta) = B(\beta) G_2 a^4 + P(\beta)$ is possible.\cite{22,23} The idea is
based on the fact that changing the number of instantons by local updating
procedure is much slower process than thermalizing the short range fluctuations
which are responsible for renormalizations. So, starting from a zero field
configuration, where $Q=0$ and $\chi=0$, and heating it will only produce
$M(\beta)$. Similarly putting one instanton of charge $Q$ on a lattice and
heating it will leave $Q$ unchanged for a large number of sweeps, but will
produce the fluctuations necessary to build up $Z$: a plateau will be reached
where $Q_L = Z Q$, and $Z$ can be read on it.

In any case it is unpleasant that most of the observed signal is artifact to be
removed. Recently, playing on the arbitrariness by higher terms in $a$ in the
definition of $Q_L$, an improved operator has been constructed for which
lattice artifacts are reduced by 2 orders of magnitude.\cite{6} $Z$ is now of
order
$0.6$, instead of $0.18$, and $M(\beta)$ is less than $10\%$ of the whole
signal~\cite{7} in a wide scaling window.

Counterterms are removed by the same non perturbative technique, but now they
are a small part of $\chi_L$. In particular $P(\beta)$ is negligible in a wide
range of $\beta$'s ($\sim 2\%$ on the entire signal).
\vskip0.05in
{\centerline{
\epsfxsize0.85\linewidth
\epsfbox{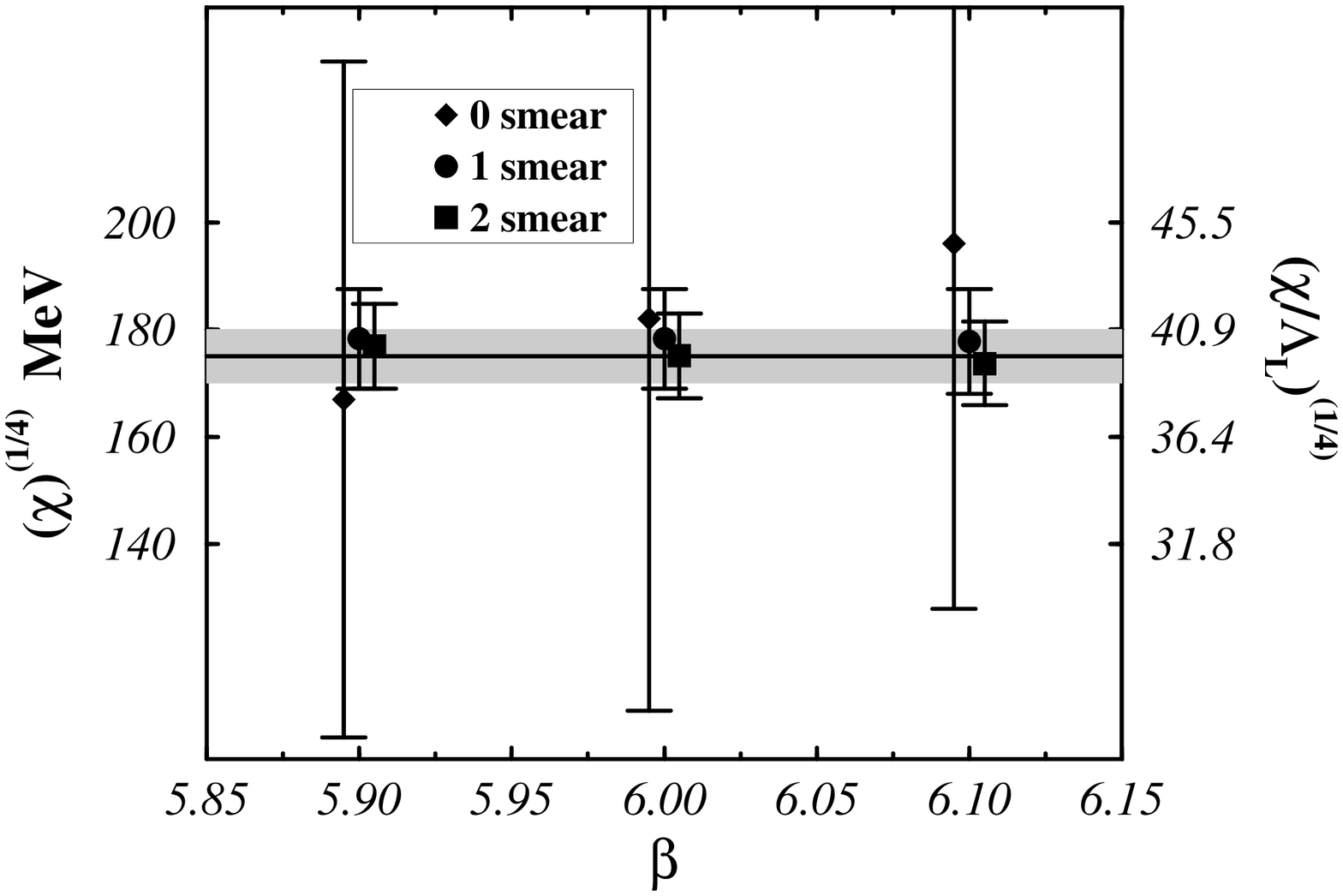}
}}
\vskip0.05in
{\centerline{
\parbox{0.9\linewidth}{
{\bf Fig.3}\,\,
Determination of $\chi$ with the naive operator, Eq.~\ref{eq:23} (0 smear) and
with the once and twice improved operators.
}
}}
\vskip0.1in
The results are summarized in Fig.~3 and in Fig.~4. Fig.~3 shows the
determination of $\chi$ at $T=0$: the result is $\chi =175\pm 5$~MeV in
excellent agreement with the Witten Veneziano prediction, and in agreement with
previous determinations~\cite{21} within their large errors. A new result is the
behaviour of
$\chi$ across the deconfining transition shown in Fig.~5, which is certainly of
interest for the models of $QCD$ vacuum based on instanton liquid.\cite{26}
\vskip0.05in
{\centerline{
\epsfxsize0.75\linewidth
\epsfbox{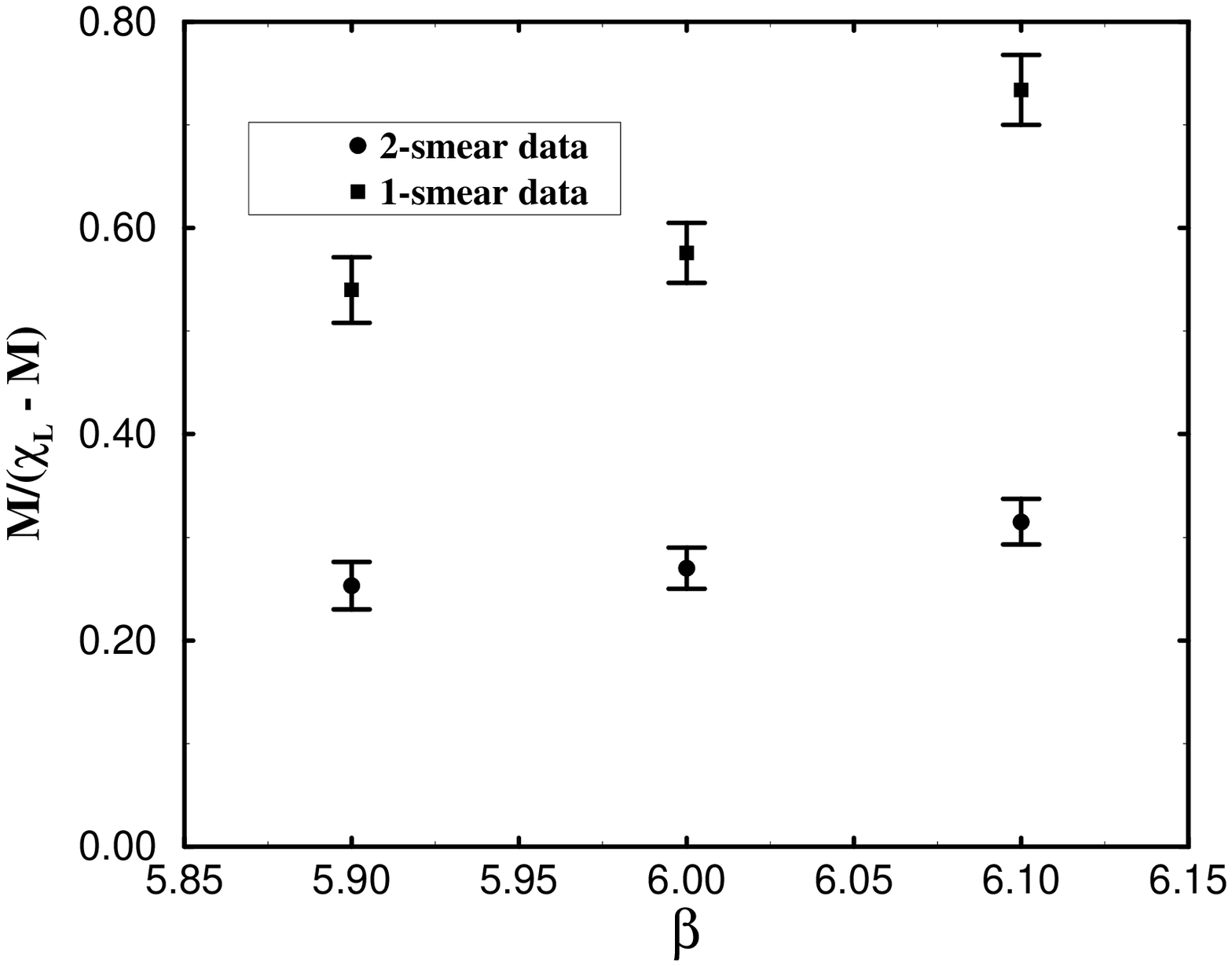}
}}
\vskip0.01in
{\centerline{
\parbox{0.95\linewidth}{
{\bf Fig.4}\,\, $M(\beta)/a^4(\beta)$. The slight deviations from
a constant allow to extract $P(\beta)$. The result is $P(\beta)< 2$-$3\%\chi_L$ 
}
}}
\vskip0.03in
{\centerline{
\epsfxsize0.70\linewidth
\epsfbox{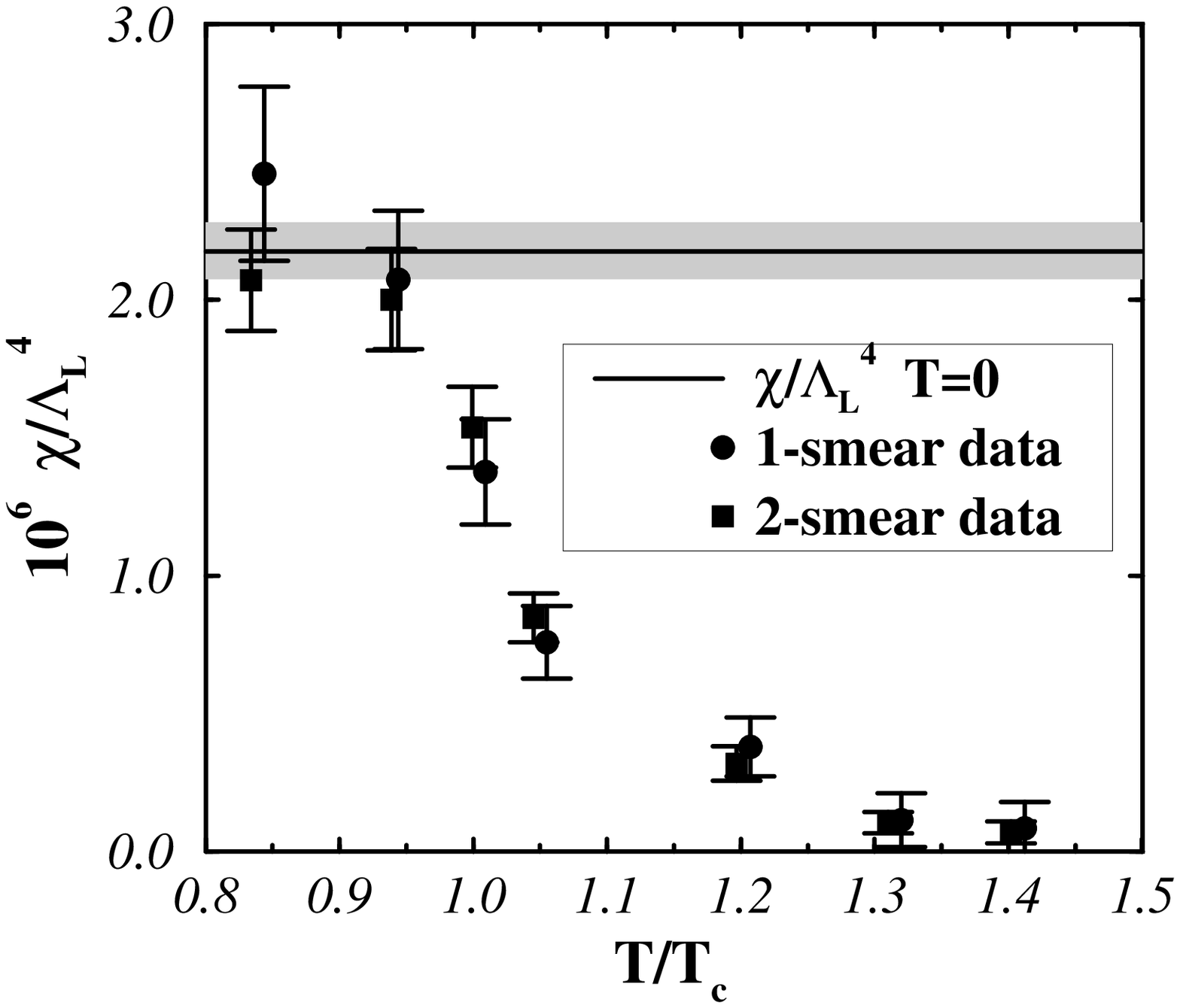}
}}
\vskip0.02in
{\centerline{{\bf Fig.5} $\chi$ across $T_c$.\,\,
}}

\vskip0.1in

\section{Gauge invariant field correlators.}
The correlators are defined  as
\begin{equation}
{\cal D}_{\mu\nu\rho\sigma}(x) =
\langle0|T\left(G_{\mu\nu}(x) S(x) G_{\rho\sigma}(0)
S^\dagger(x)\right)|0\rangle\label{eq:24}\end{equation}
with $G_{\mu\nu} = \sum_a T^a G^a_{\mu\nu}(x)$ ($T^a$ generators of the gauge
group) and $S(x)$ the parallel transport from  $0$ to $x$
\begin{equation}
S(x) = {\rm exp}\left({\rm i}\int_0^1 A_\mu(t x) x^\mu\,{\rm d} t\right)
\label{eq:25}\end{equation}
$A_\mu = \sum_a T^a A_\mu^a(x)$.

These correlators are of interest for stochastic models of confinement, where
they play a fundamental role, being the lowest order in a cluster
expansion~\cite{27,28} of the correlation functions.

The correlators had been determined on lattice a few years ago,\cite{29}
in the
range of distances from $0.4$ to $1$~fm.

Measuring ${\cal D}_{\mu\nu\rho\sigma}(x)$ on the lattice naively, taking for
$G_{\mu\nu}$ the open plaquette oerator, is difficult because of lattice
artefacts due to short range fluctuations.

If these fluctuations are smeared, e.g. by a local cooling procedure, a plateau
will eventually be reached in the cooling process, where short range effects
have been removed, but large distance physics is left unchanged. The basic idea
is that in a local cooling precess a distance $d$ is reached after a number of
steps $t$ which is governed by a sort of diffusion equation
\[ t\propto d^2\]

The minimum distance which can be explored by this technique is a few ($\sim
4$) lattice spacings.

In physical units the limitation comes from the requirement that the lattice
size $L a$ must be larger than the typical scale of $1$~fm. At $\beta\simeq
6.1$ 1~fm is 8 lattice spacings so that distances from 0.5 to 1~fm can be
explored on a $16^4$ lattice. This was the range of distances in ref.\cite{29}.
To have a distance of 3-4 lattice spacing corresponding to 0.2~fm one needs
$\beta\simeq 7$ and a lattice size $L\geq 32$. Fig.~6 shows the two independent
form factors
$D$ and
$D_1$ which parametrize~\cite{27,28} ${\cal D}_{\mu\nu\rho\sigma}$ 
as a function of distance,
extracted from the joint sample of data of ref.11 and 29.

Across $T_c$ the magnetic correlators are the same as at $T=0$, while the
electric correlators drop by an order of magnitude.\cite{31} This is consistent
with the vanishing of the string tension at $T>T_c$.

\vskip0.05in
{\centerline{
\epsfxsize0.85\linewidth
\epsfbox{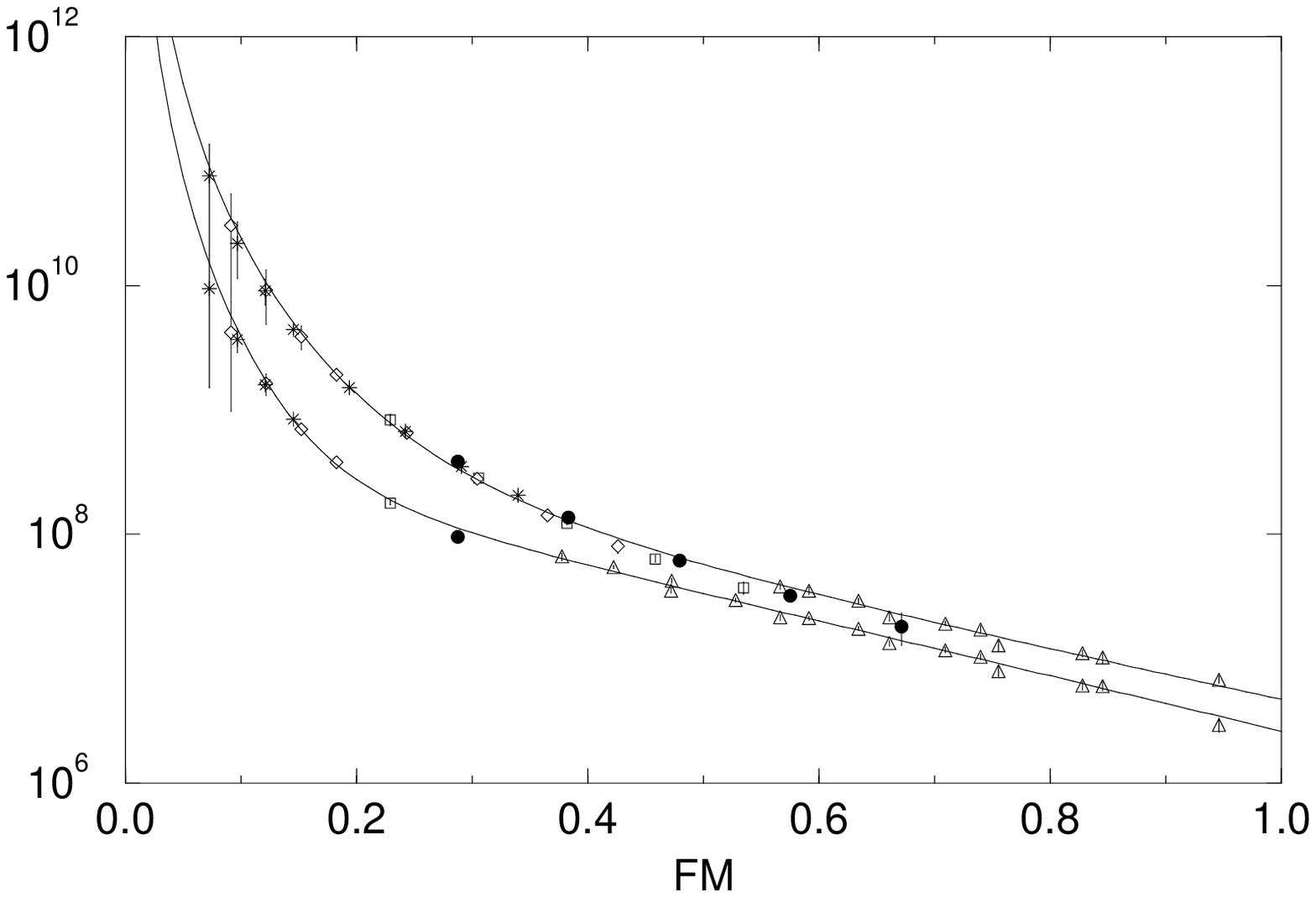}
}}
\vskip0.05in
{\centerline{
\parbox{0.9\linewidth}{
{\bf Fig.6}\,\, $D_\perp= D+D_1$ (upper curve) and 
$D_\parallel = D + D_1 + x^2\, {\partial D_1/\partial x^2}$ (lower curve).
}
}}


\vskip0.1in

\section*{References}
\begin{thebibliography}{99}
\bibitem{1}
G. 't~Hooft, in ``High Energy Physics'', EPS
International Conference, Palermo 1975, ed. A.~Zichichi;
G. 't~Hooft, {\em Nucl Phys.} {\bf B 190} (1981) 455.
\bibitem{2} S. Mandelstam, {\em Phys. Rep.} C {\bf 23}, 245 (1976).
\bibitem{3}G. Parisi, \Journal{\PRD}{11}{971}{1975}
\bibitem{8} L. Del Debbio, A. Di Giacomo, G. Paffuti,
\Journal{\PLB}{349}{513}{1995} 
\bibitem{9}L. Del Debbio, A. Di Giacomo, G. Paffuti, P. Pieri,
\Journal{\PLB}{355}{255}{1995}
\bibitem{4} E. Witten, \Journal{\NPB}{156}{269}{1979}.
\bibitem{5}G. Veneziano, \Journal{\NPB}{159}{213}{1979}.
\bibitem{6}C. Christou, A. Di Giacomo, H. Panagopoulos, E. Vicari,
\Journal{\PRD}{53}{2619}{1996}.
\bibitem{7}B. Alles, M. D'Elia, A. Di Giacomo,
{\em IFUP-TH} 26/96; hep-lat 9605013.
\bibitem{6bis}B. Alles, G. Boyd, M. D'Elia, A. Di Giacomo, E. Vicari,
{\em in preparation}.
\bibitem{29}A. Di Giacomo, H. Panagopoulos,
\Journal{\PLB}{285}{133}{1992}.
\bibitem{10}G. 't~Hooft,
\Journal{\NPB}{190}{455}{1981}.
\bibitem{11}L.P. Kadanoff, H.Ceva, 
\Journal{\PRB}{3}{3918}{1971}.
\bibitem{12}Di Cecio, A. Di Giacomo, G. Paffuti, M. Trigiante,
cond-mat/9603139
\bibitem{13}G. 't~Hooft,\Journal{\PRL}{37}{8}{1976}.
\bibitem{14}S. Weinberg,\Journal{\PRD}{11}{3583}{1975}.
\bibitem{15}G. 't~Hooft,\Journal{\NPB}{72}{461}{1974}.
\bibitem{16}G. Veneziano, \Journal{\NPB}{117}{519}{1976}
\bibitem{17}E. Witten \Journal{\NPB}{117}{269}{1979}.
\bibitem{18}G. Veneziano, \Journal{\NPB}{159}{213}{1979}.
\bibitem{19}M. Campostrini, A. Di Giacomo, H. Panagopoulos,
\Journal{\PLB}{277}{491}{1992}.
\bibitem{20}M. Campostrini, A. Di Giacomo, H. Panagopoulos, E. Vicari,
\Journal{\NPB}{329}{683}{1990}.
\bibitem{21}M. Campostrini, A. Di Giacomo, Y. G\"unduc, M.P. Lombardo,
H. Panagopoulos, R. Tripiccione,
\Journal{\PLB}{252}{436}{1990}.
\bibitem{22}A. Di Giacomo, E. Vicari,
\Journal{\PLB}{275}{429}{1992}.
\bibitem{23}B. Alles, M. Campostrini, A. Di Giacomo, Y. G\"unduc, E. Vicari,
\Journal{\PRD}{48}{2284}{1993}.
\bibitem{26}E. Shuryak, {\em Comments in Nucl. and Particle Physics},
21, 235, (1994).
\bibitem{27}H.G. Dosh, \Journal{\PLB}{190}{177}{1987}
\bibitem{28}H.G. Dosh, Yu. A. Simonov, \Journal{\PLB}{205}{339}{1988}.
\bibitem{30}A. Di Giacomo, E. Meggiolaro, H. Panagopoulos,
{\em IFUP-TH} 14/96; hep-lat 9603017
\bibitem{31}A. Di Giacomo, E. Meggiolaro, H. Panagopoulos,
{\em IFUP-TH} 14/96; hep-lat 9603018
\end{thebibliography}
\end{document}